\documentclass[%
    aip,
    jcp,
    amsmath,
    amssymb,
    twocolumn,
    superscriptaddress,
    10pt]{revtex4-1}
\usepackage{dcolumn}
\usepackage{graphicx} 
\usepackage{bm}
\usepackage{longtable}
\usepackage[english]{babel}

\begin{document}

\title{ Electrostatics of liquid interfaces } 

\author{Dmitry V.\ Matyushov} 
\email{dmitrym@asu.edu}
\affiliation{Department of Physics and Department of Chemistry \& Biochemistry, Arizona State University, PO Box 871504, Tempe, AZ 85287-1504 }

\begin{abstract}
The standard Maxwell formulation of the problem of polarized dielectrics suffers from a number of difficulties, both conceptual and practical. These difficulties are particularly significant in the case of liquid interfaces, where the ability of the interfacial multipoles to change their orientations to minimize their free energies leads to interfacial polarization localized within a thin microscopic layer. A formalism to capture this physical reality of localized interfacial polarization is proposed and is based on the surface charge as the source of microscopic electric fields in dielectrics. The surface charge density incorporates the local structure of the interface into electrostatic calculations. The corresponding surface susceptibility and interface dielectric constant provide local closures to the electrostatic boundary value problem. A robust approach to calculate the surface susceptibility from numerical simulations is proposed. The susceptibility can alternatively be extracted from a number of solution experiments, in particular those sensitive to the overall dipole moment of a closed dielectric surface. The theory is applied to the solvent-induced spectral shift and high-frequency dielectric response of solutions.     
\end{abstract}

\maketitle

\section{Introduction}
\label{sec:1}
Difficulties with continuum electrostatic models of dielectric interfaces have been recognized in the past,\cite{EygesBook:72} even though not commonly discussed. The present account aims at resolving them for interfaces in liquid dielectrics, where the problems, both conceptual and technical, are particularly difficult and pressing. The discussion starts with the outline of conceptual difficulties of Maxwell's formulation of dielectric polarization.

For a homogeneous dielectric, one commonly starts with the definition of the Maxwell electric field $\mathbf{E}$ in a continuous material made of discrete molecules carrying molecular charges. Those are typically represented by distributed partial charges of atoms and molecular groups. We adopt this convention here and consider the permanent charge distribution and neglect the electronic molecular polarizability. One, therefore, only needs to deal with the changes in positions and orientations of the molecules, leading to fluctuating internal electric fields. These fields are strong and highly non-uniform on the length-scale of individual molecules. The standard approach is to smooth out the variations of the microscopic internal field $\bm{\mathcal{E}}$ over a ``physically small volume'' $\Omega$: $\mathbf{E}_p= \langle  \bm{\mathcal{E}} \rangle_{\Omega}$. The dimensions of the volume $\Omega$ need to be small relative to the length-scale of a particular measurement, but large enough to contain many molecules.\cite{Landau8} The Maxwell field is then defined as the sum of the field $\mathbf{E}_0$ of external charges and the smoothed-out field of internal charges 
\begin{equation}
\mathbf{E}=\mathbf{E}_0+\mathbf{E}_p. 
\label{eq:00}
\end{equation}
The measurable quantity is actually not the field itself, but its (macroscopic) line integral defining the voltage difference $V=\int \mathbf{E}\cdot d\mathbf{l}$. The standard dielectric experiment measures the material's dielectric constant $\epsilon$ as the drop of the voltage in the material compared to vacuum.\cite{EygesBook:72,Purcell:63} 

The macroscopic dielectric constant $\epsilon$ also enters the Maxwell constitutive relation connecting the field $\mathbf{E}$ to the displacement vector $\mathbf{D}$
\begin{equation}
\mathbf{D} = \epsilon \mathbf{E} .
\label{eq:0}
\end{equation}
The electric field and displacement vectors enter the free energy density of the electric field in the dielectric as conjugate variables,\cite{Landau8} $\propto \mathbf{D} \cdot \mathbf{E}$. The displacement vector thus bears an analogy with the displacement of a physical body under the action of a force, associated with the electric field, such that the mechanical work is given by their scalar product.  

The constitutive relation is supplemented by boundary conditions at dielectric interfaces. The tangential component of the longitudinal (see below) electric field is continuous at the dielectric interface, while the transverse displacement vector preserves its normal component. These boundary conditions, together with the constitutive relation given by Eq.\ \eqref{eq:0}, complete the boundary value problem of the Maxwell electrostatics.\cite{EygesBook:72,Landau8,Purcell:63,Jackson:99} 

The uniform macroscopic electric field $E=V/d$ is directly accessible from the dielectric experiment measuring the voltage $V$ across the plates separation $d$. On the contrary,  how to define the generally inhomogeneous field $\mathbf{E}_p$ in Eq.\ \eqref{eq:00} at the micro-to-meso length scale has never been adequately resolved. It might appear to have become a straightforward task with the advent of numerical simulations, but exactly how one should perform the average $\langle\dots\rangle_{\Omega}$ remains unclear. The fields typically reported in the bulk materials by either simulations or by spectroscopy are those produced at a given target molecule by the surrounding condensed phase. It is, however, well established that this local field  (which is often identified with the cavity field\cite{Onsager:36}) is distinct from $\mathbf{E}$. In fact, the connection between the two fields has been sought by essentially all mean-field theories of dielectrics.\cite{Onsager:36,Boettcher:73} Alternatively, following the idea originally advanced by Kelvin for magnetic materials\cite{Thompson1872} and by Maxwell for dielectrics,\cite{Maxwell:V2} one can measure the field inside a hollow cavity in the dielectric. However, this approach inevitably requires an interface and the corresponding interfacial polarization when external fields are applied. We address the problem of the field inside a cavity in our discussion below because of its close relation to the general issue of defining fields inside materials, even though the dielectric constant can be defined without invoking cavities.\cite{Fulton:1975ga} What needs to be stressed though is that only the voltage difference, local field at a target particle, or a field next to an interface can be measured experimentally.  

The discussion presented below starts with the general properties of fields in dielectrics based on the Helmholtz theorem.\cite{EygesBook:72,Jackson:99} We then proceed to the formulation of the boundary value problem and the formalism of extracting the dielectric susceptibility of the interface from numerical simulations, followed by the connection of the theory to spectroscopic and dielectric experiments.

\section{Longitudinal and transverse fields}

\begin{figure}
\includegraphics*[width=5cm]{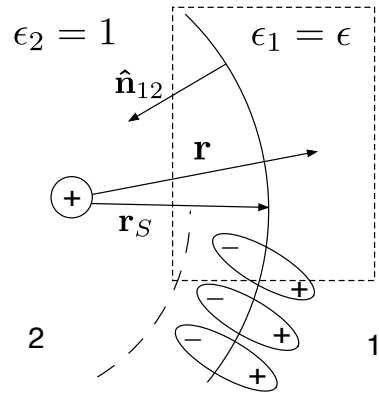}
\caption{Cartoon of the interface between a dielectric with the dielectric constant $\epsilon$ and vacuum. The source of external field is indicated by a positive point charge. The dividing, dielectric-vacuum surface is shown by the solid line. It cuts through surface molecules of the dielectric producing a surface charge density. The latter is sensitive to the choice of the dividing surface since a surface drawn inside the void (long dashed line) produces no surface charge. The area indicated by the dashed rectangle separates a part of the interface with no external charges; 
$\mathbf{\hat n}_{12}$ is the surface normal pointing from medium 1 to medium 2.  
}
\label{fig:1}
\end{figure}

Introduction of interfaces into dielectrics makes conceptual difficulties more severe. In order to set up the problem, we will consider an interface between vacuum and a dielectric polarized by some external charges indicated by a positive point charge in Fig.\ \ref{fig:1}. We will next consider a part of the interface where there are no external charges, shown by the dashed rectangle in the figure. 

The first conceptual problem appears in introducing a dividing surface between dielectrics of different polarity. One can draw a mathematical surface separating the dielectric from a void. This infinitely thin mathematical surface will cut through some surface molecules, remove the corresponding molecular charges, and create the surface charge density\cite{Jackson:99,Landau8} 
\begin{equation}
\label{eq:1}
\sigma(\mathbf{r}_S) = \mathbf{P}(\mathbf{r}_S)\cdot\mathbf{\hat n} .
\end{equation}
Here, $\mathbf{P}$ is the dipolar polarization density of the dielectric and the normal unit vector $\mathbf{\hat n}$ is directed outward to the dielectric; $\mathbf{r}_S$ is the position at the surface. It is immediately clear that the concept of the dividing surface, and the corresponding surface charge, even though a purely macroscopic construct, requires recognizing the molecular granularity of the material and the separation of charge within the molecule. A surface drawn within the void (dashed line in Fig.\ \ref{fig:1}) will produce zero surface charge and thus will not capture the interfacial polarization by external fields. There is a clear conceptual contradiction between the macroscopic character of the dividing surface and the microscopic distribution of molecular charge and the orientational molecular order at the interface on which the surface charge density must depend. 

The exact position of the surface inside the dielectric does not need to be well defined when fields are uniform. In that case, the total dipole between two surfaces of arbitrary shape is zero, and the calculations are not affected by the surface position.\cite{Frohlich} This is obviously not true for inhomogeneous fields, as is well documented when dielectric cavities need to be defined in solvation models. We next show that inhomogeneous fields present even more severe conceptual difficulties since Eq.\ \eqref{eq:0} connects fields of different symmetry, longitudinal for $\mathbf{E}$ and transverse for $\mathbf{D}$.  

Since no external charges are present in the region within the selected area in Fig.\ \ref{fig:1}, the first differential Maxwell equation for the displacement vector $\mathbf{D}$ reads  $\nabla\cdot \mathbf{D}=0$ for any point $\mathbf{r}$ within the  region. The displacement vector connects the Maxwell field $\mathbf{E}$ to the dipolar polarization density $\mathbf{P}$ as follows 
\begin{equation}
\label{eq:2}
4\pi\mathbf{P}=\mathbf{D} -\mathbf{E} .
\end{equation} 
The displacement vector is transversal  (divergence-free, $\nabla\cdot \mathbf{D}=0$), while the electric field is longitudinal  (curl-free, $\mathbf{E}=-\nabla\Phi$, $\Phi$ is the electrostatic potential). Therefore, Eq.\ \eqref{eq:2} represents, according to the Helmholtz theorem,\cite{EygesBook:72} the separation of the polarization field into longitudinal ($L$) and transverse ($T$) components.\cite{Felderhof:1977do,Kivelson:89} 

The Helmholtz theorem\cite{EygesBook:72,Jackson:99} is a general mathematical statement stipulating that any inhomogeneous vector field $\mathbf{A}$ can be split into the longitudinal (irrotational)  $\mathbf{A}_L$ and transverse (solenoidal) $\mathbf{A}_T$ components such that $\nabla\cdot \mathbf{A}_T=0$, $\nabla\times \mathbf{A}_L=0$, and $\int \mathbf{A}_L\cdot \mathbf{A}_T d\mathbf{r}=0$.  The longitudinal component  is
\begin{equation}
\mathbf{A}_L = -\frac{1}{4\pi} \nabla \int \frac{\nabla' \mathbf{A}'}{|\mathbf{r}-\mathbf{r}'|} d\mathbf{r}' ,  
\label{eq:2-1}
\end{equation}
where here and below $\mathbf{A}'=\mathbf{A}(\mathbf{r}')$. When this equation is applied to a general vector field of the dipolar polarization $\mathbf{P}$,  
one directly gets
\begin{equation}
4\pi\mathbf{P}_L = - \mathbf{E}_p,
\label{eq:2-2}
\end{equation}
where $\mathbf{E}_p=-\nabla\Phi_p$ and $\Phi_p$ is the electrostatic potential created by the molecular charges of the dielectric  
\begin{equation}
\Phi_p = \int \frac{\rho_p'}{|\mathbf{r}-\mathbf{r}'|} d\mathbf{r}' .
\label{eq:3-1}
\end{equation}
Here, $\rho_p=-\nabla\cdot \mathbf{P}$ is the polarization charge density. 

Equation \eqref{eq:2-2} indicates that the knowledge of the longitudinal component of the polarization density gives access to the field $\mathbf{E}_p$ of the molecular charges. It also implies $\mathbf{E}_p=0$ outside of the dielectric (or inside a void) where $\mathbf{P}_L=0$. The remaining transverse component of the polarization density contributes to the displacement vector. Specifically, since $\nabla\cdot\mathbf{E}_0=0$  ($\mathbf{E}_0=-\nabla\Phi_0$) in the selected region, one gets  
\begin{equation}
4\pi\mathbf{P}_T = \mathbf{D} - \mathbf{E}_0. 
\label{eq:3-2}
\end{equation}
Combining Eqs.\ \eqref{eq:2}, \eqref{eq:2-2}, and \eqref{eq:3-2}, the Maxwell field becomes
\begin{equation}
\mathbf{E}=-\nabla \Phi,\quad \Phi = \Phi_0+\Phi_p . 
\label{eq:3}
\end{equation}

Equation \eqref{eq:3} indicates that the electric field inside the material is the sum of the field of the volume polarization and the external field. This result is of course consistent with the microscopic picture of fields within materials, which, according to the Coulomb law, are caused by the combined effect of the external  and internal charges. In the presence of interfaces, the polarization field $\mathbf{P}$ is highly inhomogeneous, on the molecular length-scale, in the interface. The macroscopic polarization, averaged over a physically small region $\Omega$, does not account for these effects and the microscopic molecular structure of the interface is required to calculate the electrostatic potential arising from the interface. Given the difficulty of reliable calculation of interfacial $\mathbf{P}$, we will coarse-grain the interfacial molecular structure into the surface charge density $\sigma(\mathbf{r}_S)$. 
This property requires introducing the mathematical dividing surface, which itself carries conceptual difficulties specified above. Therefore, a formalism to calculate the surface charge density needs to be additionally supplied. This is achieved below by relating  $\sigma(\mathbf{r}_S)$ to either correlation functions involving the interfacial dipolar polarization provided by numerical atomistic simulations or to experimental observables. We first turn to conceptual difficulties arising when applying the constitutive Maxwell relation given by Eq.\ \eqref{eq:0} to the electrostatics of dielectric interfaces.  

\section{Maxwell dielectrics}
Solving differential Maxwell equations requires a constitutive relation connecting $\mathbf{D}$ and $\mathbf{E}$. For isotropic materials, this is commonly supplied in the form of a simple proportionality given by Eq.\ \eqref{eq:0}. The equation $\nabla\cdot\mathbf{D}=0$ then reduces to the Laplace equation $\Delta \Phi =0$ in a piecewise homogeneous medium. This equation is supplemented by the boundary conditions 
\begin{equation}
\epsilon_1\mathbf{\hat n}_{12}\cdot\nabla\Phi_{1}= \epsilon_2\mathbf{\hat n}_{12}\cdot\nabla\Phi_{2} 
\label{eq:3-0}
\end{equation}
and $\Phi_1=\Phi_2$ at the dividing surface. Here, $\mathbf{\hat n}_{12}$ is the unit normal to the surface directed from region 1 to region 2; $\epsilon_1$ and $\epsilon_2$ are the dielectric constants of the materials at contact. 

The Maxwell constitutive relation also implies the proportionality between $\mathbf{P}$ and $\mathbf{E}$ through the susceptibility $\chi=(\epsilon-1)/(4\pi)$, $\mathbf{P}=\chi\mathbf{E}$. One therefore obtains\cite{EygesBook:72}
\begin{equation}
\nabla\cdot\mathbf{P}=\nabla\cdot\mathbf{P}_L=\chi\left(\nabla\cdot\mathbf{E}_0+\nabla\cdot\mathbf{E}_p\right) .
\label{eq:3-3}
\end{equation}
On the other hand, from Eq.\ \eqref{eq:2-2},
\begin{equation}
\nabla\cdot\mathbf{P}_L=-4\pi \nabla\cdot \mathbf{E}_p .
\label{eq:3-4}
\end{equation}
Since $\nabla\cdot\mathbf{E}_0=0$ in the interfacial region, Eqs.\ \eqref{eq:3-3} and \eqref{eq:3-4} can be simultaneously satisfied only when $\nabla\cdot \mathbf{P}=0$. This requirement implies that the volume charge density inside the dielectric is zero, 
$\rho_p=0$, and from Eq.\ \eqref{eq:3-1} 
\begin{equation}
\Phi_p=0.
\label{eq:3-5}
\end{equation}

Deriving Eq.\ \eqref{eq:3-5} requires only the Helmholtz theorem, the transverse character of $\mathbf{D}$ in the dielectric, and  the constitutive Maxwell equation [Eq.\ \eqref{eq:0}]. Its surprising result is the disappearance of the potential of the bulk charges $\Phi_p$ in the overall Maxwell potential in Eq.\ \eqref{eq:3}. For microscopic dielectrics, this result indicates that the Maxwell constitutive relation cannot be correct in the interface where the polarization field experiences fast variations. For the continuum representation of the dielectric, the missing term is of course the potential created by the surface charge, which appears when the microscopic interface is replaced by the mathematical dividing surface. The surface charge density $\sigma(\mathbf{r}_S)$ is given by Eq.\ \eqref{eq:1}. In the Maxwell formulation, the polarization density in Eq.\ \eqref{eq:1} is related to the potential $\Phi$ through the bulk dielectric susceptibility $\chi$. We adopt below an alternative approach in which the surface charge density, incorporating microscopic properties of the interface, is supplied as the constitutive relation. The condition of disappearing volume polarization charge, $\rho_p=0$, is used to get rid of the  polarization potential $\Phi_p$. The problem of the dielectric response can then be consistently formulated based solely on the Coulomb law.

\section{Electrostatic boundary problem}
\label{sec:2}
We start with the microscopic polarization density 
\begin{equation}
\mathbf{P}(\mathbf{r}) =  \sum_j \langle\mathbf{m}_j \delta\left(
                           \mathbf{r}-\mathbf{r}_j\right)\rangle  
\label{eq:4}
\end{equation}
specified by the positions $\mathbf{r}_j$ and orientations of the medium dipoles $\mathbf{m}_j$; angular brackets denote an ensemble average. The microscopic electrostatic potential $\Phi_m$ directly follows from the dipole-truncated multipolar expansion of the Coulomb law 
\begin{equation}
\label{eq:5}
\Phi_m = \Phi_0 + \int \nabla'\frac{1}{|\mathbf{r}-\mathbf{r}'|} \cdot \mathbf{P}' d\mathbf{r}' .
\end{equation}
By using the Gauss theorem, this equation transforms into
\begin{equation}
\label{eq:6}
\Phi_m = \Phi_0 + \Phi_p + \Phi_S,
\end{equation}
where $\Phi_p$ is the scalar potential given by Eq.\ \eqref{eq:3-1} and the last summand is the potential of the surface charge
\begin{equation}
\label{eq:7}
\Phi_S = \oint\frac{\sigma(\mathbf{r}_S) dS}{ |\mathbf{r}-\mathbf{r}_S|} ,
\end{equation}
which appears in the overall potential as a result of introducing the dividing surface. As mentioned above, there is no dividing surface for a microscopic interface and polarization decays continuously into the void. The surface integral does not appear in Eq.\ \eqref{eq:6}, which is an exact consequence of the Helmholtz theorem. It is clear that the surface term needs to be ``introduced'', by providing a dividing surface, into the electrostatic potential and does not necessarily follow from the general properties of inhomogeneous vector fields used to describe dielectrics.     

The microscopic electric field $\mathbf{E}_m = -\nabla \Phi_m$ is a non-local property, which is the fundamental reason for the difficulties with Maxwell's constitutive relations. It is given by the equation
\begin{equation}
\mathbf{E}_m = \mathbf{E}_0 +\int \mathbf{T}(\mathbf{r}-\mathbf{r}')\cdot \mathbf{P}'d\mathbf{r}',
\label{eq:8}
\end{equation}
in which $\mathbf{T}(\mathbf{r}-\mathbf{r}')=-\nabla\nabla' \left|\mathbf{r}-\mathbf{r}'\right|^{-1}$ is the long-range dipolar tensor combining both longitudinal and transverse components and thus propagating the corresponding components of $\mathbf{P}$. The real-space convolution in Eq.\ \eqref{eq:8} is eliminated only in inverted $\mathbf{k}$-space, where this equation becomes an algebraic relation.  

One can next use the Helmholtz theorem [Eq.\ \eqref{eq:2-1}] to simplify Eq.\ \eqref{eq:8}. The result differs inside and outside of the dielectric. Since $4\pi\mathbf{P}_L=-\mathbf{E}_p=0$ outside the dielectric, the microscopic field becomes
\begin{equation}
\mathbf{E}_m=\mathbf{E}_0 + \mathbf{E}_S,
\label{eq:8-1}
\end{equation}
where $\mathbf{E}_S= -\nabla \Phi_S$. When the field  is calculated inside the dielectric, a small region around the point of singularity of $\mathbf{T}(\mathbf{r}-\mathbf{r}')$ needs to be taken out,\cite{Jackson:99} with the result
\begin{equation}
\mathbf{E}_m=\mathbf{E}_0-\frac{8\pi}{3}\mathbf{P}_L+\frac{4\pi}{3}\mathbf{P}_T +  \mathbf{E}_S .
\label{eq:8-2}
\end{equation}
When the Maxwell constitutive relation is used in Eq.\ \eqref{eq:8-2}, one gets
\begin{equation}
\mathbf{E}_m=\frac{\epsilon+2}{3} \mathbf{E} +  \mathbf{E}_S .
\label{eq:8-3}
\end{equation}
The first summand here is the Lorentz field,\cite{Boettcher:73} which is the only term required for $\mathbf{E}_m$ far from interfaces where $\mathbf{E}_S$ vanishes. 
 
Both equations \eqref{eq:8-1} and \eqref{eq:8-2} show that the only non-local part of the microscopic field $\mathbf{E}_m$ caused by the polarized dielectric comes from the field of the surface charges. According to Eq.\ \eqref{eq:8-1}, it is the only field of the polarized dielectric that a measuring device (e.g., a spectroscopic probe) placed either outside of the dielectric or inside a void can directly detect. It is, therefore, this field, and the corresponding surface charge density $\sigma$, that is the main focus of our formalism. 

The approach we adopt here is to put $\Phi_p=0$ in Eq.\ \eqref{eq:6}, following the discussion leading to Eq.\ \eqref{eq:3-5}. Here we follow Eyges,\cite{EygesBook:72,Eyges:1975wa} who applied this anzatz to dielectrics in general to eliminate contradictions of the standard Maxwell formulation. In this approach, all polarization of the dielectric body contributing to $\Phi_m$ is concentrated in the interface, also in agreement with the standard dielectric experiment in which $\mathbf{P}=Const$ and $\rho_p=0$.  

While this approach is just a convenient approximation for dielectrics in general, it provides the correct physical picture for liquid dielectrics. The interfacial dipoles of liquids can nearly freely change their orientation to minimize the surface free energy.\cite{Frenkel:55} Effective screening of the perturbation produced by creating the interface occurs as the result of this structural adjustment, and both the density and orientational perturbations of the liquid propagate only a few molecular layers into the bulk.\cite{Sokhan:97,Faraudo:2004fk,DMpre1:08,Pershan:12,Gekle:2012kx,Horvath:2013fe} The polar response is then dominated by the interface, and the language of interfacial polarization is the most relevant for describing polarized polar liquids.\cite{DMpre1:08} If, next, the microscopic orientational structure of the interface is incorporated into the definition of the surface charge density, one can arrive at a physically motivated formulation of the electrostatics of liquid dielectrics. This is the program of this development, which also attempts to identify experimental observables probing local properties of the interface in order to connect them to the surface charge density. 

The potential $\Phi_m = \Phi_0+\Phi_S$ is created by external and surface charge sources and satisfies the Poisson equation. In order to formulate the boundary conditions, one recalls that the normal component of the field should be discontinuous at the dividing surface, with the discontinuity related to the surface charge density\cite{Jackson:99}
\begin{equation}
\label{eq:9}
\mathbf{\hat n}_{12}\cdot\nabla\Phi_{m1}= \mathbf{\hat n}_{12}\cdot\nabla\Phi_{m2} +
4\pi \sigma ,
\end{equation}
where the surface charge density is given as
\begin{equation}
\label{eq:10}
\sigma=(\mathbf{P}_1-\mathbf{P}_2)\cdot \mathbf{\hat n}_{12} .
\end{equation}
We next proceed to identifying the constitutive relations connecting the surface charge density to the electric field $\mathbf{E}_m=-\nabla \Phi_m$.

\section{Surface dipolar susceptibility}
\label{sec:3}
The potential of polarized charges $\Phi_m$ is a solution of the Poisson equation, satisfying continuity of $\Phi_m$  at the dividing surface and the second boundary condition given by Eqs.\ \eqref{eq:9} and \eqref{eq:10}. The second boundary condition needs to be closed by relating $\sigma$ to $\Phi_m$ or $\mathbf{E}_m$. This connection is achieved in the plane capacitor dielectric experiment. The electric field inside the capacitor is obviously $E=E_m=4\pi(\sigma_\text{ext}+\sigma)$, where $\sigma_\text{ext}$ is the surface charge density at the capacitor's plates. Measuring the capacitance at constant charge with and without the dielectric specifies the susceptibility linking $\sigma$ to $\sigma_\text{ext}$: $\sigma=\chi_{\sigma}\sigma_\text{ext}$, $\chi_{\sigma}=\epsilon^{-1}-1<0$. The problem of calculating the potential $\Phi_p$ is avoided in this experimental setup by the condition $\mathbf{P}=Const$ and $\rho_p=0$. 

Following the logic of the plane capacitor calculation, one needs to find the
susceptibility connecting $\sigma$ to $\mathbf{E}_0$. This can be achieved by using the linear response theory.\cite{Hansen:03,Felderhof:1977do} To simplify the discussion, we  consider the dividing surface separating the dielectric from a void, as in Fig.\ \ref{fig:1}.  The interface between two dielectrics follows from subtracting two dielectric/void solutions or, alternatively, by replacing $\epsilon$ with $\epsilon_1/\epsilon_2$ since only the ratio of the dielectric constants enters the boundary value problem.\cite{Landau8}
 
The projection of the polarization density field on the surface normal $\mathbf{\hat n}_{12}$ can be calculated in the linear response approximation under the common assumption of a weak external field $\mathbf{E}_0$
\begin{equation}
\label{eq:11}
\langle P_n\rangle =\beta 
\int \langle\delta P_n \delta \mathbf{P}_L'\rangle \cdot \mathbf{E}_0' d\mathbf{r}' ,
\end{equation}
where $P_n=\mathbf{\hat n}_{12}\cdot\mathbf{P}(\mathbf{r}_S)$, $\beta=1/(k_\text{B}T)$ is the inverse temperature, and the angular brackets denote an ensemble average. Since $\mathbf{E}_0$ is a longitudinal field, only the longitudinal polarization density $\mathbf{P}_L'$ gives a nonzero contribution to the integral over $\mathbf{r}'$.  The longitudinal and transverse fields mostly fluctuate independently.\cite{Felderhof:1977do,DMjcp3:08} Therefore, both the surface polarization $\mathbf{P}(\mathbf{r}_S)$ and the correlation function between the polarization fluctuations $\delta \mathbf{P}$ in Eq.\ \eqref{eq:11} refer to their longitudinal projections. For a continuum dielectric  $\beta\langle\delta P_{L,\alpha}(\mathbf{r})\delta P_{L,\beta}(\mathbf{r}')\rangle = \chi_L\delta_{\alpha\beta}\delta(\mathbf{r}-\mathbf{r}')$, where $\chi_L=(1-\epsilon^{-1})/(4\pi)$ is the longitudinal dielectric susceptibility and $\alpha,\beta$ are the Cartesian components of the vector fields.\cite{Madden:84}  

If the range of the external field variation exceeds the correlation length of the polarization density, the external field can be taken out of the integral in Eq.\ \eqref{eq:11}. If one additionally neglects correlations between normal and tangential projections of the polarization density, one gets
\begin{equation}
\label{eq:12}
\langle P_n\rangle =\chi_{0n}E_{0n},\quad\chi_{0n}=\beta \langle\delta P_n \delta M_{sn} \rangle ,
\end{equation}
where  $E_{0n}=\mathbf{\hat n}_{12}\cdot\mathbf{E}_0(\mathbf{r}_S)$ and
\begin{equation}
\label{eq:13}
\mathbf{M}_s=\int \mathbf{P}_L(\mathbf{r}) d\mathbf{r} 
\end{equation}
is the total dipole moment of the solvent. 

\begin{figure}
\includegraphics*[clip=true,trim= 0cm 1.5cm 0cm 0cm, width=7cm]{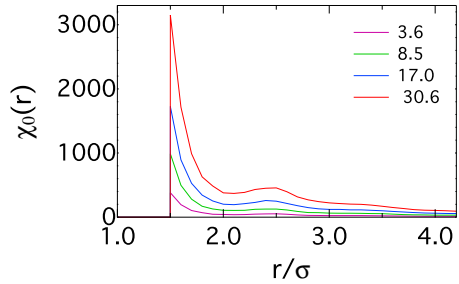}
\caption{Susceptibility $\chi_0(r)$ from Eq.\ \eqref{eq:13-1} calculated for spherical concentric layers around the spherical cavity of the radius $R_0/\sigma=1.0$ in fluids of dipolar hard spheres. The distance from the cavity's center is scaled with the hard-sphere diameter $\sigma$. Dielectric constants of dipolar fluids are shown in the legend and the fluids' density is $\rho\sigma^3=0.8$. }
\label{fig:2}
\end{figure}

The requirement to use longitudinal fields in  Eq.\ \eqref{eq:12} makes this relation largely impractical for the direct analysis of simulations. As is seen from Eqs.\ \eqref{eq:2-1}--\eqref{eq:3-1} the calculation of the longitudinal projection $\mathbf{P}_L$ from the overall polarization density $\mathbf{P}$, directly available from simulations, requires convoluting the polarization density of the entire simulation box with the dipolar tensor at each instantaneous configuration. This calculation needs to be repeated for each point $\mathbf{r}$ where the longitudinal polarization density is calculated. Failing to limit the consideration by longitudinal fluctuations incorporates strong transverse polarization fluctuations, which do not couple to the longitudinal electric field and cannot contribute to the susceptibility.

In order to illustrate the extent of error introduced by transverse fluctuations, we show in Fig.\ \ref{fig:2} the susceptibility calculated 
from the overall polarization density of a spherical layer, correlated with the total dipole moment $\mathbf{M}_s$ of the simulation cell. The polarization density is calculated for radial layers of the fluid of dipolar hard spheres around a spherical cavity. The corresponding distance-dependent susceptibility is given by the following relation  
\begin{equation}
\chi_0(r) = \beta \langle \mathbf{P}(r)\cdot\mathbf{M}_s \rangle .
\label{eq:13-1}
\end{equation}
The details of the simulation protocol and the data analysis are given in the Supplementary Material (SM),\cite{supplJCP} here we discuss only the results.

\begin{figure}
\includegraphics*[width=7cm]{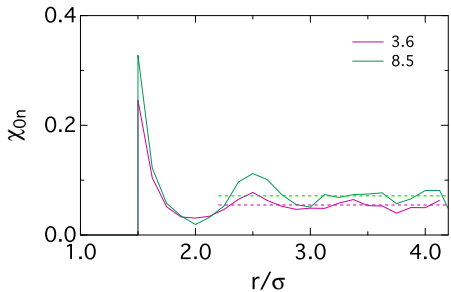}
\caption{$\chi_{0n}(r)$ from Eq.\ \eqref{eq:13-2} calculated for concentric layers around a spherical cavity of the radius $R_0/ \sigma =1.0$ in dipolar hard-sphere fluids with dielectric constants indicated in the plot. The dashed lines indicate the results of calculations using a linear fit of the radial correlation functions in Eq.\ \eqref{eq:13-3} (see text and SM\cite{supplJCP}). All system parameters are the same as in Fig.\ \ref{fig:2}, the dielectric constants of dipolar fluids are indicated in the legend.  
}
\label{fig:3}
\end{figure}

The susceptibility $\chi_0(r)$ spikes to very high values near the interface, then slowly decays to the bulk. Qualitatively similar results have been reported for aqueous interfaces.\cite{DMcpl:11,Bonthuis:2012mi,Gekle:2012kx}  On the other hand, 
the longitudinal symmetry of the polarization field, which makes direct calculations according to Eq.\ \eqref{eq:12} largely impractical, can be used to obtain an equation more suitable for numerical applications. By applying Eqs.\ \eqref{eq:2-2} and \eqref{eq:3-1} one can re-write Eq.\ \eqref{eq:11} as follows
\begin{equation}
\langle P_n\rangle =\beta\sum_i q_i \langle \delta P_n \delta \Phi_{p,i} \rangle .
\label{eq:13-2}
\end{equation}
Here, the sum is over the external charges $q_i$ and fluctuations $\delta \Phi_{p,i}$ of the solvent potential at those charges. If the source of the electrostatic field is the solute dissolved in a liquid, the above equation can be re-written as
\begin{equation}
\langle P_n\rangle =\beta \langle \delta P_n \delta E^{\text{C}}\rangle,
\label{eq:13-2}
\end{equation}
where $E^{\text{C}}$ is the solute-solvent electrostatic (Coulomb) interaction energy. This relation directly provides a map of $\sigma(\mathbf{r}_S)$ of the solvation layer surrounding a solute.  

Figure \ref{fig:3} illustrates the application of this formula when one probe charge is placed at the center of the spherical cavity carved from the fluid of dipolar hard spheres. The susceptibility is calculated for concentric shells of radius $r$ around the cavity and thus becomes a function of $r$. As is seen, eliminating the transverse fluctuations from $\chi_0(r)$ in Eqs.\ \eqref{eq:12} and \eqref{eq:13-2} significantly reduces the susceptibility. It still preserves its spike at the distance of the closest approach of the solvent to the cavity and oscillations decaying into the bulk. 

The uncertainty with the choice of the interfacial susceptibility $\chi_{0n}$ from the distance-dependent, and oscillatory, function $\chi_{0n}(r)$ can be resolved by calculating the integrated radial dipole moment of the hydration layer within the $r$-shell: $M(r)=\sum_{r_j<r} \mathbf{m}_j\cdot \mathbf{\hat r}_j$, $\mathbf{\hat r}_j=\mathbf{r}_j/r_j$. For a charge placed at the center of the cavity one gets
\begin{equation}
\chi_{0n}(r) = (\beta/4\pi)\frac{d}{dr}\langle \delta M(r) \delta\Phi_p(0) \rangle ,
\label{eq:13-3}
\end{equation}
where $\Phi_p(0)$ is the solvent potential at the center of the cavity. As is shown in the SM,\cite{supplJCP}  $\langle \delta M(r) \delta\Phi_p(0) \rangle$ is well represented by a linear function of $r$, thus producing a constant $\chi_{0n}$ for the derivative in Eq.\ \eqref{eq:13-3}. The results of these calculations are shown by the dashed lines in Fig.\ \ref{fig:3}, confirming that taking the derivative in Eq.\ \eqref{eq:13-3} is consistent with averaging the oscillations of $\chi_{0n}(r)$ out. We turn to the polarity of the interface below again, but first discuss the closure of the boundary value problem in Eqs.\ \eqref{eq:9} and \eqref{eq:10} in terms of the electrostatic potential $\Phi_m$ and define the dielectric constant of the interface.

We will use the plane capacitor geometry to establish the connection between $\chi_{0n}$ and the susceptibility to the overall field $\langle P_n\rangle =\chi_{n}E_{n}$, $E_{n}=\mathbf{\hat n}_{12}\cdot\mathbf{E}_m(\mathbf{r}_S)$. For the plane capacitor, $E_{n}=E_{0n}-4\pi \langle P_n\rangle$ (note the convention for the surface normal), and one gets
\begin{equation}
\label{eq:14}
\chi_{n}=\chi_{0n}/\left(1-4\pi\chi_{0n}\right) .
\end{equation}    
The second boundary condition can now be re-written in the form commonly accepted in the theories of dielectrics
\begin{equation}
\tilde\epsilon_1\mathbf{\hat n}_{12}\cdot\nabla\Phi_{m1}= \tilde\epsilon_2\mathbf{\hat n}_{12}\cdot\nabla\Phi_{m2} ,
\label{eq:15}
\end{equation}
where 
\begin{equation}
\tilde\epsilon_i =1+4\pi\chi_{n,i} = \left(1-4\pi\chi_{0n,i}\right)^{-1} .
\label{eq:16}
\end{equation}

Spontaneous polarization is possible when dipole and quadrupole moments of the liquid compete to minimize their free energy in the interface.\cite{Frenkel:55} This effect is particularly strong for water,\cite{Horvath:2013fe} where the competition is between an axial dipole and a mostly non-axial quadrupole. The spontaneous polarization of the interface leads, according to Eq.\ \eqref{eq:1}, to a spontaneous surface charge density $\sigma_s$, which exists even at zero field. The total surface charge density then combines the response to the electric field with the spontaneous component: $\sigma=\sigma_s+\chi_nE_n$. Equation \eqref{eq:15} modifies to
\begin{equation}
\tilde\epsilon_1\mathbf{\hat n}_{12}\cdot\nabla\Phi_{m1}= \tilde\epsilon_2\mathbf{\hat n}_{12}\cdot\nabla\Phi_{m2} +4\pi\sigma_s . 
\label{eq:15-1}
\end{equation}
In addition to a non-zero offset of the surface charge density, one can also anticipate a possibility of $\chi_n$ depending on the sign of $E_n$ to reflect the well-established asymmetry of the water's linear response to ions of opposite charge.\cite{Hummer:96,Rajamani:04} This dependence of susceptibility on ion's charge is prominent for small ions, but disappears with the growing size of the solute once the charge-dipole interaction becomes comparable with $k_\text{B}T$.\cite{Ashbaugh:00}

The equations for the potential $\Phi_m$ are the same as Maxwell's equations and, therefore, standard numerical Poisson equation solvers can be used. Similarly to the Maxwell formulation, the theory requires only one susceptibility parameter. On the other hand, even though Eq.\ \eqref{eq:15-1} casts the problem of interfacial electrostatics in the familiar terms of the boundary value problem, the knowledge of the susceptibility $\chi_{0n}$ responding to the field of external charges is sufficient for the direct calculation of the electrostatic potential $\Phi_m$ from the known distribution  of external charges and the corresponding electric field $\mathbf{E}_0$. Although we cannot prove it here, the connection between $\chi_n$ and $\chi_{0n}$ given by Eq.\ \eqref{eq:14} might be nonuniversal (the universality of the corresponding relation for bulk dielectrics is simply an experimental fact). The formulation in terms of $\mathbf{E}_0$ and $\chi_{0n}$ is preferable from this perspective.

The ``interface dielectric constant'' $\tilde\epsilon_i$ in Eqs.\ \eqref{eq:16} and \eqref{eq:15-1} will in most cases of interest be distinct from the standard dielectric constant supplied by the dielectric experiment, which is specified by the ``tilde'' sign.  However, for the dielectric polarization in a plane capacitor we get $\tilde\epsilon =\epsilon $ and the standard longitudinal susceptibility $\chi_{0n}=\chi_L=\left(1-\epsilon^{-1}\right)/(4\pi)$. Therefore, for a plane capacitor, $\tilde\epsilon$  yields the enhancement of the capacitance reported experimentally as the dielectric constant of the material. This agreement is not expected to hold for more complex interfacial geometries and nonuniform external fields, as we show next for the problem of spherical cavities in fluids of dipolar hard spheres.

\begin{figure}
\includegraphics*[clip=true,trim= 0cm 1.5cm 0cm 0cm,width=7cm]{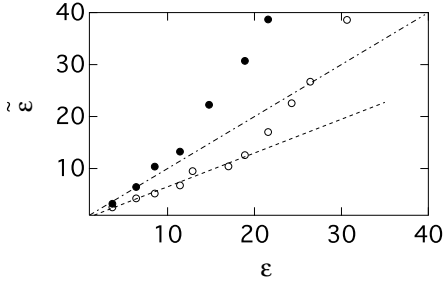}
\caption{$\tilde\epsilon$ calculated from Eqs.\ \eqref{eq:16} and \eqref{eq:13-3} vs.\ the bulk $\epsilon$ of fluids of dipolar hard spheres. The points are the simulation results for the cavity radius $R_0/\sigma=1.0$ (closed circles) and $R_0/\sigma=3.0$ (open circles). The dashed-dotted line marks $\tilde\epsilon=\epsilon$ to guide the eye. The dashed line with the slope 0.65 is a linear regression through seven lowest polarity points shown by the open circles. }  
\label{fig:4}
\end{figure}

The free energy of polarizing the dielectric can be written in terms of the polarization charges as follows\cite{Landau8}
\begin{equation}
\Delta F = \tfrac{1}{2}\int \sigma \Phi_0 dS + \tfrac{1}{2}\int \rho_p \Phi_0 d\mathbf{r}.  
\label{eq:61}
\end{equation}
In the current model, $\rho_p=0$ and only the first integral appears in the free energy. On the contrary, for a microscopic interface, there is no dividing surface and only the second integral contributes. Since the polarization free energy should not depend on the model, one can use this condition to formulate the sum rule for $\sigma$. By applying Eq.\ \eqref{eq:2-2}, we can write this condition in the form
\begin{equation}
\int \sigma \Phi_0 dS = \sum q_i\Phi_{p,i} ,
\label{eq:62}
\end{equation}
where, as in Eq.\ \eqref{eq:13-2} above, $q_i$ are the external charges producing the external potential $\Phi_0$; $\Phi_{p,i}$ are the potentials of the polarized solvent at the positions of these charges. In the specific case of a single ion at the center of a spherical void of radius $a$ one gets
\begin{equation}
\Phi_p(0) = 4\pi a\sigma_0 = - 4\pi a P_r\big|_{r=r_S},  
\label{eq:63}
\end{equation}
where $P_r$ is the radial projection of the average polarization density and $\sigma_0$ is the angular-averaged surface charge density ($\ell=0$ expansion term in Eq.\ \eqref{eq:17} below). The potential $\Phi_p(0)$ at the void's center adds to the  potential difference at the planar liquid-air interface to make the electrochemical potential measuring the work of transferring an ion from the gas phase into its cavity inside the liquid.\cite{Kathmann:2011vn,Beck:2013gp,Horvath:2013fe}    

We illustrate the application of the sum rules in Eq.\ \eqref{eq:62} to numerical simulations of cavities in dipolar fluids.\cite{DMpre1:08} We consider polarization of the dielectric by an ion with charge $q$ placed at the center of the cavity of radius $a$. Dielectric models suggest that the product of the surface charge density with the surface area $S=4\pi a^2$ remains constant 
\begin{equation}
- q^{-1}\sigma_0 S = 1- \epsilon^{-1} ,
\label{eq:64}
\end{equation}
This relation, used in the sum rule in Eq.\ \eqref{eq:63}, yields $\Phi_p(0)=-(1-\epsilon^{-1})(q/a)$ and the corresponding Born solvation free energy $(q/2)\Phi_p(0)$. 

\begin{figure}
\includegraphics*[clip=true,trim= 0cm 1.5cm 0cm 0cm,width=7cm]{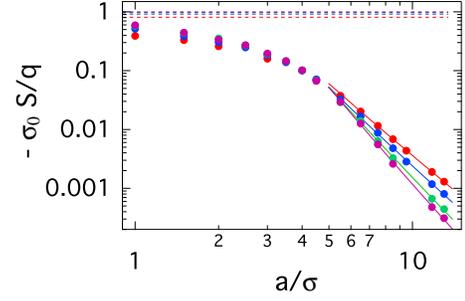}
\caption{$-\sigma_0S/q$ vs.\ $a/\sigma=R_0/\sigma+1/2$ for cavities in dipolar hard spheres with $\epsilon=3.6$ (red), 8.5 (blue), 30.6 (green), and 93.7 (magenta); $R_0$ is the radius of the hard-sphere cavity and $R_0+\sigma/2$ is the distance of the closest approach of the solvent. The solid lines are linear fits to large cavity portions of the data. The regression slopes are: $-3.8$ (red), $-4.1$ (blue), $-4.7$ (green), and $-5.1$ (magenta). The dashed lines indicate the results of Eq.\ \eqref{eq:64}. }
\label{fig:5}
\end{figure}

The potential $\Phi_p(0)$ can be extracted from simulations of cavities in liquids by the use of the linear response approximation,\cite{Hansen:03} which suggests that the average potential of the solvent in response to the charge can be obtained from the variance of the potential fluctuations in the absence of the charge.\cite{DMpre1:08} One then gets
\begin{equation}
-q^{-1}\sigma_0 S = (\beta a/2) \langle (\delta\Phi_p(0))^2\rangle .
\label{eq:65}
\end{equation}     
The results of MC simulations\cite{DMpre1:08} for three fluids of dipolar hard spheres surrounding cavities of varying radius are shown in Fig.\ \ref{fig:5}. The points are the simulation results plotted against the cavity size. They are compared to the predictions of Eq.\ \eqref{eq:64} shown by the dashed lines. 

The results presented in Fig.\ \ref{fig:5} illustrate why the focus on the interfacial properties is required for a physically motivated model of liquid electrostatics. As the size of the cavity grows, the orientational structure of the interface changes, thus altering the corresponding surface charge density and the cavity potential related to it. As a result of these structural changes, the invariance of $\sigma_0 S$ suggested by Eq.\ \eqref{eq:64} does not hold anymore. A new scaling $\sigma_0S \propto a^{-\alpha}$, $\alpha\simeq 4-5$ emerges, which is not anticipated by the standard electrostatic arguments. With this new scaling, the solvent response $\Phi_p(0)$ to a charge inside a void scales down faster than $a^{-1}$ for voids larger than the critical size approximately four times the size the solvent molecule. We note that the average potential $\Phi_p(0)$ may contain a contribution $\Phi_s(0)$ arising from the spontaneous polarization of the interface, which is zero for dipolar liquids\cite{Horvath:2013fe} considered here. For water, $\Phi_s(0)$ saturates to a nearly constant positive value at $a\simeq 12-17$ \AA\ and needs to be subtracted from the overall $\Phi_p(0)$ to address the issue of scaling with the cavity size.\cite{Ashbaugh:00} 

The appearance of a faster than $a^{-1}$ decay of the cavity potential can be explained from the following arguments. In the standard picture, charge $q$ placed in the center of a void generates the radial polarization $P_r=\chi q/(\epsilon r^2)$ propagating into the volume of the dielectric. The solvent potential at the position of the charge is then
\begin{equation}
\Phi_p(0)=-\int_{r>a} (P_r/r^2) d\mathbf{r}\propto a^{-1} 
\end{equation}
If, on the contrary, the liquid polarization is limited to a thin interfacial layer, the above integral can be written as
\begin{equation}
\Phi_p(0) = -\frac{\chi q}{\epsilon a^4} N_d(a) v_d,
\label{eq:50}
\end{equation}
where the interface is represented by $N_d(a)$ correlated domains with the domain volume $v_d$. If dipoles respond independently, $N_d(a)\propto a^2$ and one expects the $a^{-2}$ scaling for $\Phi_p(0)$. This is not observed in simulations, suggesting that the dipolar response of the interface is both localized and highly correlated. The origin of the $\alpha\simeq 4-5$ exponent in the decay of $\sigma_0S$ with the radius $a$ is currently not clear. However, the direct physical consequence of this result is that internal charges of large solutes receive less solvation stabilization than is traditionally expected. Solubility of large solutes requires surface charges and corresponding surface solvation.\cite{DMjpcl2:12}

\section{Polarized cavity in a uniform electric field}
\label{sec:3}
\begin{figure}
\includegraphics*[width=4.5cm]{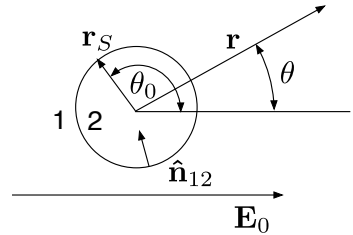}
\caption{Spherical cavity polarized by the field of external charges $\mathbf{E}_0$.}
\label{fig:6}
\end{figure}

Here we illustrate the new boundary conditions for the electrostatics of liquids by applying them to the problem of a  spherical void polarized by a uniform external field (Fig.\ \ref{fig:6}). This problem directly applies to the high-frequency dielectric response of solutions\cite{DMjpcm:12} and to optical spectroscopy as we discuss below.

In order to come up with specific parameters of the void's interface, we will use the axial symmetry of the problem and expand $\sigma$ in Legendre polynomials $P_{\ell}(\cos\theta_0)$ of the polar angle $\theta_0$ between the $z$-axis aligned with the external field and the position $\mathbf{r}_S$ at the dividing spherical surface (Fig.\ \ref{fig:6})
\begin{equation}
\label{eq:17}
\sigma(\theta_0) = \sum_{\ell} \sigma_{\ell} P_{\ell}(\cos\theta_0) .
\end{equation}
It is easy to see that expansion terms of order $\ell$ are connected to the $z$-components of surface multipoles of the corresponding order. One gets the $z$-projection of the interfacial dipole $M_{0z}^{\text{int}}=\sigma_1\Omega_0$ at $\ell=1$ and $zz$-projection $Q_{0,zz}^{\text{int}}=(3/5)\sigma_2 a\Omega_0$ of the surface quadrupole at $\ell=2$; $\Omega_0$ is the void's volume.

The potential $\Phi_m$ (see SM\cite{supplJCP}) is the solution of the Poisson equation for a spherical cavity polarized by a uniform external electric field such that the local field $\mathbf{E}_m$ becomes the Maxwell field $\mathbf{E}$ in the bulk of the polarized dielectric. From this solution, the local electrostatic field at the void's center is 
\begin{equation}
\label{eq:18}
E_{m,z}(0)  = E - (4\pi /3) \sigma_1 . 
\end{equation}
The field at the center of the cavity is fully defined by the interface dipole, or the $\sigma_1$ projection of the surface charge density. 

One can make one step further to connect $\sigma_1$ to $\chi_n$. The field $E_{n}$ at the cavity's surface (SM\cite{supplJCP}), is related, through $\chi_n$, to the surface charge density given by Eq.\ \eqref{eq:17}. By equating the $\ell=1$ components in the two equations, one gets the desired relation
\begin{equation}
\sigma_1 = - \frac{\chi_n E}{1+ (8\pi/3)\chi_n}  = - \frac{\chi_{0n} E}{1 - (4\pi/3)\chi_{0n}} .
\label{eq:19}
\end{equation}
In terms of the local dielectric constant $\tilde\epsilon$ in Eq.\ \eqref{eq:16}, this relation becomes
\begin{equation}
\sigma_1= -\frac{3}{4\pi}\ \frac{\tilde\epsilon -1}{2\tilde\epsilon+1} E.
\label{eq:20}
\end{equation}
The second term in this equation is easily recognized as the dielectric reaction field response appearing in  theories dielectrics;\cite{Onsager:36,Boettcher:73} $E_{m,z}(0)$ in Eq.\ \eqref{eq:18} then becomes the corresponding expression for the cavity field.\cite{Boettcher:73} If one assumes $\tilde\epsilon=\epsilon$, one arrives at the Maxwell result for the dipole moment induced at a spherical void by a uniform external field\cite{Jackson:99}
\begin{equation}
M_{0z}^{\text{int}} = - \frac{3\Omega_0}{2\epsilon+1}P,
\label{eq:21}
\end{equation}
where $P=(4\pi)^{-1}(\epsilon-1)E$ is the dielectric polarization far away from the interface. 
 
These calculations illustrate that the results of Maxwell's electrostatics are a part of the proposed formalism. However, the polarization of the interface of a polar liquid facing a void deviates significantly from the predictions of Maxwell's electrostatics due to anisotropic orientational structure of the interfacial dipoles, implying that Eq.\ \eqref{eq:21} does not agree with simulations,\cite{DMjcp2:11} or, alternatively, $\tilde\epsilon\ne\epsilon$. The current formulation provides more flexibility to account for such results by connecting electrostatic calculations to experiments reporting local interfacial properties, such as interfacial multipolar moments.

\section{Connection to experiment}
Electrostatic fields in condensed media are traditionally quantified by the solvent induced shift of optical\cite{Mataga:70} or vibrational\cite{Levinson:2012ly} transition lines. The electrostatic component of the shift, often dominant,\cite{Reynolds:96} is given by the product of the dipole moment change of the chromophore $\Delta m$ and the reaction field $R$ in the ground state in equilibrium with the ground-state dipole $m_g$
\begin{equation}
h\Delta\nu = - \Delta m R .
\label{eq:25}
\end{equation}

From derivations presented above, $R=-(4\pi/3)\sigma_1$, but the interface dipole, represented by $\sigma_1$, now needs to be updated with the account for the polarizing field of the central dipole of the chromophore $m_g$. Repeating the steps leading to Eq.\ \eqref{eq:20}, we obtain
\begin{equation}
\sigma_1 = -\frac{2m_g}{a^3}\ \frac{\chi_n}{1+(8\pi/3)\chi_n} . 
\label{eq:26}
\end{equation}
From this relation, the reaction field becomes
\begin{equation}
R=\frac{m_g}{a^3}\ \frac{(8\pi/3)\chi_n}{1+(8\pi/3)\chi_n} = \frac{2m_g}{a^3}\ \frac{\tilde\epsilon -1}{2\tilde\epsilon+1}.
\label{eq:27}
\end{equation}
One recovers the standard Onsager relation\cite{Onsager:36} for $R(\epsilon)$ when $\tilde\epsilon=\epsilon$. Note that $R(\epsilon)-R(\epsilon_{\infty})$ is often used in Eq.\ \eqref{eq:25} to separate the response due to the high-frequency dielectric constant $\epsilon_{\infty}$ from the total polar response $R(\epsilon)$. This is not required in Eq.\ \eqref{eq:27} since $\chi_n$ can be understood as the permanent dipole susceptibility of the interface. This susceptibility can, therefore, be tabulated by spectroscopic shifts of dipolar dyes.\cite{Reynolds:96} The same function can be extracted from high-frequency dielectric measurements of solutions as we discuss next.  

When the frequency of the capacitor field or of radiation exceeds the characteristic relaxation frequency of the solute dipole, the solution response approaches that of the solution of voids. The measurement of the absorption coefficient or the dielectric constant of the solution gives access to the dipolar susceptibility of the cavities produced by the excluded solute volumes in the solvent, $\chi_1=\sigma_1/E$.\cite{DMjpcm:12} The solution dielectric constant $\epsilon_{\text{sol}}$ can be found from the equation\cite{DMjpcm:12} 
\begin{equation}
\label{eq:28}
(\epsilon/\epsilon_{\text{sol}}) = 1+\eta_0(\epsilon-1)\left[1+(8\pi/3)\chi_1 \right] 
\end{equation}
in which $\eta_0$ is the solute volume fraction. In this equation, both $\epsilon$ and $\epsilon_\text{sol}$ should be understood as $\epsilon(\omega)$ and $\epsilon_{\text{sol}}(\omega)$ at $\omega_0\ll \omega\ll \omega_D$, where $\omega_0^{-1}$ is the relaxation time of the solute dipole and $\omega_D^{-1}$ is the characteristic (Debye) relaxation time of the solvent.  

The slope of the dielectric decrement $\Delta \epsilon=\epsilon_{\text{sol}}-\epsilon$ vs.\ $\eta_0$ gives access to the susceptibility $\chi_1=\sigma_1/E$ and, through Eq.\ \eqref{eq:19}, to $\chi_n$. Experimental absorption data for aqueous solutions show that the slope, and $\chi_1$ extracted from it, change significantly depending on the nature of the solute and the corresponding interfacial structure of water.\cite{DMjcp3:12} Dielectric constant obviously does not capture these variations.  Therefore, measurements of solution absorption can potentially replace the dielectric experiment in providing the local interface susceptibility. 

Equations \eqref{eq:25}, \eqref{eq:27}, and \eqref{eq:28} suggest that high-frequency dielectric and spectroscopic measurements give access to the same interface susceptibility. One can, therefore, combine these equations into a 
relation including only experimentally accessible properties of solutions 
\begin{equation}
\label{eq:29}
(\epsilon/\epsilon_{\text{sol}}) = 1+\eta_0(\epsilon-1)\left[1+h\Delta\nu a^3/(\Delta m m_g)\right] .
\end{equation} 

An alternative approach to access the susceptibility $\chi_1$ is through dielectrophoresis. Electric field gradient exerts a force on the interface dipole $M_{0z}^\text{int}$. The force along the $z$-direction is given by the expression\cite{JonesBook:95}
\begin{equation}
F_{0z}=\frac{3\Omega_0\epsilon}{8\pi} K \frac{\partial}{\partial z} E^2, 
\label{eq:28-1}
\end{equation}
where $K$ is the dielectrophoresis constant. Since the free energy of a polarized (non-polar) solute is equal to $-M_{0z}^\text{int} E_0$ it follows from the previous discussion that 
\begin{equation}
K = (4\pi/3)\chi_1. 
\label{eq:28-2}
\end{equation}
The dielectric constant of an ideal solution gives, therefore, access to the dielectrophoresis constant and, alternatively, dielectrophoresis of non-polar particles can be used as input to predict the dielectric constant of an ideal solution in Eq.\ \eqref{eq:28}. We note that dielectric models give the following expression for $K$\cite{JonesBook:95}
\begin{equation}
K=\frac{\epsilon_2-\epsilon_1}{\epsilon_2+2\epsilon_1} , 
\label{eq:29}
\end{equation}
where we explicitly specified the dielectric constant of the solute $\epsilon_2$ (Fig.\ \ref{fig:1}). This equation can be contrasted with Eq.\ \eqref{eq:28-2} to quantify surface polarization effects not accounted for by the standard formulation. 

While mobility in the field gradient gives access to the interface dipole susceptibility $\chi_1$, measuring mobility of particles in a uniform electric field (electrophoresis\cite{Hunter:01}) provide access to the quadrupole moment of the interface, $Q_{0,zz}^{\text{int}}=(3/5)\sigma_2a\Omega_0$. This second multipolar moment of the interface arises from a non-zero second-order term $\sigma_2$ in the expansion of the surface charge density in rotational invariants [Eq.\ \eqref{eq:17}]. \cite{DMmp:14} The anisotropic orientational structure of the interface creates conditions for an anisotropic response to the applied field. The difference in response results in different values of the local electric field $E_m$ on the opposite sides of the suspended particle and, correspondingly, different electrostatic energy densities. This condition results in a gradient of the chemical potential projected on a dragging force acting on the suspended particle. The force along the external field applied to a particle carrying the charge $q$ becomes\cite{DMmp:14}  
\begin{equation}
F_z = \left(q + 2Q_{0,zz}^{\text{int}}/a^2 \right) E .
\label{eq:30}
\end{equation}
Measuring the force, or electrophoretic mobility, provides experimental access to the quadrupole moment of a closed interface. 

\section{Conclusions}
\label{sec:5}
The Maxwell constitutive relations $\mathbf{D}\propto \mathbf{E}$, $\mathbf{P}\propto \mathbf{E}$ establish simple proportionality rules between fields of different symmetry. The longitudinal and transverse components of the polarization density field $\mathbf{P}$, mixed in the Maxwell constitutive relations, carry distinctly different physical properties.\cite{Kivelson:89} This distinction is particularly pronounced when looking at the $k\to 0$ behavior of their corresponding structure factors:\cite{Madden:84} nearly flat for the longitudinal projection and strongly peaked, and infinitely increasing at the ferroelectric transition, for the transverse projection.\cite{DMjcp3:08} Since continuum electrostatics is recovered in the $k\to 0$ limit of $k$-dependent dipolar response functions,\cite{Dolgov:81} there is a good physical reason for the linear constitutive relations to be successful for the longitudinal projection, but they are expected to fail for the transverse projection. The distinctions between longitudinal and transverse projections of the polarization field, well established for homogeneous dielectrics, require particular care when considering interfaces. 

The enormous simplification provided by the constitutive relations is the ability to cast the problem in terms of the Poisson equation with a single susceptibility (dielectric constant) entering the boundary conditions.\cite{Jackson:99} This procedure presents some clear conceptual difficulties, but, from the practical perspective, has also run into problems when applied to molecular-size objects\cite{Fedorov:2007fa,DMjcp3:08} and to nanometer-scale liquid interfaces.\cite{DMpre1:08,DMjcp2:11} The deviations from the expected behavior are not limited to quantitative disagreements in calculated electrostatic energies, but reach the level of qualitative differences. The scaling of the liquid polar response to an ion placed in the center of a void\cite{DMpre1:08} shows a cross-over from the expected $\propto a^{-1}$ scaling (Born model) to $\propto a^{-(4-6)}$ scaling with increasing void's radius $a$ (Fig.\ \ref{fig:5}). In addition, the electric field inside a void carved from a uniformly polarized dielectric does not reach the value $\simeq (3/2)E$ at $\epsilon\gg 1$,\cite{Onsager:36,Boettcher:73} but instead shows the behavior consistent with the Lorentz formula for the cavity field $\simeq (\epsilon/3) E$.\cite{DMjcp3:08,DMjcp2:11} 

These effects, and potentially a number of others, are different manifestations of the same physical phenomenon: the localization of the polar response in the liquid's interfacial layer, instead of its spreading through the dielectric, as anticipated by the Maxwell picture\cite{Faraudo:2004fk} (see discussion leading to Eq.\ \eqref{eq:50}). For the localized interfacial response, the specific orientational structure of the interfacial multipoles strongly affects the electrostatic potential and field produced by the interface. The electrostatic problem can then be recast in terms of the interface-specific surface charge density.

An approach consistent with the picture of surface polarization dominating the electrostatic response of a liquid dielectric\cite{DMpre1:08,DMjcp3:08,DMjcp2:11} is proposed here. It reformulates the boundary value electrostatic problem in terms of the surface charge density and the corresponding surface charge susceptibility. Importantly, the local surface susceptibility is introduced for the longitudinal component of the polarization field only, thus avoiding transverse polarization fluctuations strongly enhancing the susceptibility, but decoupled from the longitudinal external field. This formalism offers a robust route to the calculation of the dielectric constant of the interface.  

The interface susceptibility characterizes specific interfaces and aims to replace the dielectric constant as the input into the electrostatic boundary value problem. It can be calculated from polarization correlation functions supplied by numerical simulations or related to experimental observables. In particular, the dipole moment of the interface (the first multipole of the surface charge density) enters a number of observables characterizing solutions polarized by long-wavelength stationary or oscillatory (e.g., radiation) fields.

\acknowledgments This research was supported by the National Science
Foundation (CHE-1213288). CPU time was provided by the National Science Foundation through XSEDE resources (TG-MCB080116N). The author is grateful to Daniel Martin for his help with the data analysis.

\bibliography{chem_abbr,dielectric,dm,statmech,protein,liquids,solvation,dynamics,elastic,simulations,surface,nano,water,ir,et,glass}

\end{document}